\documentclass[onecolumn,preprintnumbers,amssymb,amsmath,superscriptaddress]{revtex4}[10pt]
\usepackage{axodraw}
\usepackage{pstricks}
\usepackage{color}
\usepackage{graphicx}
\usepackage{dcolumn}
\usepackage{bm}

\begin{document} 
\newcommand{\beq}{\begin{eqnarray}}
\newcommand{\eeq}{\end{eqnarray}}
\newcommand{\nn}{\nonumber}
\def\ltap{\ \raise.3ex\hbox{$<$\kern-.75em\lower1ex\hbox{$\sim$}}\ }
\def\gtap{\ \raise.3ex\hbox{$>$\kern-.75em\lower1ex\hbox{$\sim$}}\ }
\def\CO{{\cal O}}
\def\CL{{\cal L}}
\def\CM{{\cal M}}
\def\tr{{\rm\ Tr}}
\def\CO{{\cal O}}
\def\CL{{\cal L}}
\def\CM{{\cal M}}
\def\mpl{M_{\rm Pl}}
\newcommand{\bel}[1]{\be\label{#1}}
\def\al{\alpha}
\def\bt{\beta}
\def\eps{\epsilon}
\def\eg{{\it e.g.}}
\def\ie{{\it i.e.}}
\def\mn{{\mu\nu}}
\newcommand{\rep}[1]{{\bf #1}}
\def\be{\begin{equation}}
\def\ee{\end{equation}}
\def\bea{\begin{eqnarray}}
\def\eea{\end{eqnarray}}
\newcommand{\eref}[1]{(\ref{#1})}
\newcommand{\Eref}[1]{Eq.~(\ref{#1})}
\newcommand{\gsim}{ \mathop{}_{\textstyle \sim}^{\textstyle >} }
\newcommand{\lsim}{ \mathop{}_{\textstyle \sim}^{\textstyle <} }
\newcommand{\vev}[1]{ \left\langle {#1} \right\rangle }
\newcommand{\bra}[1]{ \langle {#1} | }
\newcommand{\ket}[1]{ | {#1} \rangle }
\newcommand{\ev}{{\rm eV}}
\newcommand{\kev}{{\rm keV}}
\newcommand{\Mev}{{\rm MeV}}
\newcommand{\gev}{{\rm GeV}}
\newcommand{\tev}{{\rm TeV}}
\newcommand{\mev}{{\rm MeV}}
\newcommand{\meV}{{\rm meV}}
\newcommand{\mnu}{\ensuremath{m_\nu}}
\newcommand{\nnu}{\ensuremath{n_\nu}}
\newcommand{\mlr}{\ensuremath{m_{lr}}}
\newcommand{\acc}{\ensuremath{{\cal A}}}
\newcommand{\mav}{MaVaNs}
\newcommand{\disc}[1]{{\bf #1}} 
\newcommand{\mh}{{m_h}}
\newcommand{\hb}{{\cal \bar H}}
\newcommand{\me}{\mbox{${\rm \not\! E}$}}
\newcommand{\met}{\mbox{${\rm \not\! E}_{\rm T}$}}
\newcommand{\MPl}{M_{\rm Pl}}
\newcommand{\ra}{\rightarrow}
\pagestyle{plain}

\title{Flavor in Supersymmetry with an Extended $R$-symmetry}
\author{Graham D. Kribs}
\affiliation{\mbox{Department of Physics and Institute of Theoretical Science, University of Oregon, Eugene, OR 97403, USA}}
\author{Erich Poppitz}
\affiliation{Department of Physics, University of Toronto, 60 St.\ George St., Toronto, ON M4Y 3B6, Canada}
\author{Neal Weiner}
\affiliation{Center for Cosmology and Particle Physics,
  Department of Physics, New York University,
New York, NY 10003, USA}
\preprint{}
\date{\today}
\begin{abstract}

We propose a new solution to the supersymmetric flavor problem
without flavor-blind mediation.  Our proposal is to enforce 
a continuous or  a suitably large discrete $R$-symmetry on weak scale supersymmetry, so that
Majorana gaugino masses, trilinear $A$-terms, and the $\mu$-term are 
forbidden.  We find that replacing the MSSM with an $R$-symmetric supersymmetric model 
allows order one flavor-violating soft masses, even for squarks 
of order a few hundred GeV.  The minimal $R$-symmetric
supersymmetric model contains Dirac gaugino masses and $R$-symmetric 
Higgsino masses with no left-right mixing in the squark or slepton
sector.  Dirac gaugino masses of order a few TeV with vanishing $A$-terms 
solve most flavor problems, while the $R$-symmetric Higgs sector becomes 
important at large $\tan\beta$.  $\epsilon_K$ can be accommodated if CP is preserved in the SUSY breaking sector, or if there is a moderate flavor degeneracy, which can arise naturally. $\epsilon'/\epsilon$, as well as neutron and electron EDMs are easily within experimental bounds. The most striking phenomenological distinction of this
model is the order one flavor violation in the squark and slepton sector,
while the Dirac gaugino masses tend to be significantly heavier than
the corresponding squark and slepton masses.

\end{abstract}

\maketitle
\twocolumngrid

\section{Introduction} 
With the LHC soon to commence, attention has increasingly turned to the question of what signals one might expect to see. Within the context of a variety of new models, specifically supersymmetry, little Higgs theories, and theories with new, TeV-scale dimensions, there has been a broad phenomenology already described.

Up to this point, however, there has been a ubiquitous feature regarding flavor. In  theories of physics beyond the standard model, especially with light states which carry standard model flavor quantum numbers, it has been generally found that flavor violation must be extremely suppressed, in particular in the lighter two generations. This can be understood either in terms of effective flavor-changing operators, or, within the context of a particular theory such as supersymmetry, in terms of explicit flavor violating spurions \cite{Gabbiani:1988rb,Gabbiani:1996hi,Bagger:1997gg,Ciuchini:1998ix}.

There are a number of flavor violating observables which constrain such new physics: $K$--$\bar{K}$ oscillations, $b\rightarrow s \gamma$, $B_s \rightarrow \mu^+ \mu^-$, $B \rightarrow \tau \nu$, $\Delta M_B$, $\Delta M_{Bs}$, $\mu \rightarrow e \gamma$, $\tau \rightarrow \mu \gamma$, and $\mu$-$e$ conversion, to name several. Of these, $K$--$\bar{K}$ is typically the most constraining, in terms of the size of flavor violation, because it is so suppressed in the standard model. For instance, for $500$~GeV squarks in the MSSM, with gluinos of a similarly ``natural'' size of $500$~GeV, the off diagonal elements---as usual, taken as dimensionless ratios  $\delta_{ij}= (m^2_{\tilde{q}})_{ij}/|m_{\tilde{q}}^2|$---of the squark mass-squared matrices  must obey $\delta_{LL} < .06$ in the best case scenario that $\delta_{RR,LR}=0$, and $\sqrt{\delta_{LL}\delta_{RR}}  < 10^{-3}$, $\sqrt{\delta_{LR}\delta_{RL}} < 2 \times 10^{-3}$ under more general assumptions.

Such limits apparently instruct us that whatever mediates supersymmetry breaking to the observable sector, it should be flavor blind. This has inspired a great deal of work on mediation mechanisms that are sufficiently flavor diagonal, such as gauge mediation \cite{Dine:1993yw,Dine:1994vc,Dine:1995ag} (see  \cite{Giudice:1998bp} for a review), anomaly mediation \cite{Randall:1998uk,Giudice:1998xp}, or gaugino mediation \cite{Kaplan:1999ac,Chacko:1999mi}. Alternative proposals \cite{Cohen:1996vb} are to push the lighter two generations above $m_{\tilde q} \sim 50$~TeV ($600$~TeV if CP is maximally violated or $5-20$~TeV with moderate flavor degeneracy) where the flavor violation would not effect precision observables, but, unfortunately, would not be detectable by the LHC either.

An exciting possibility would be that there is significant flavor violation in new physics, but the nature of the new physics ``screens'' it sufficiently from the existing observables. In this paper, we consider such a possibility within the context of supersymmetry that contains an extended $R$-symmetry (i.e., an $R$-symmetry larger than $R$-parity alone).  We will generally take the extended $R$-symmetry to be continuous, even though a $Z_4$ $R$-symmetry is sufficient for practically all purposes.  Moreover, for brevity we will refer to our ``extended $R$-symmetry'' as simply ``the $R$-symmetry'' throughout the paper.

The layout of this paper is as follows.  In Sec.~\ref{sec:rsym} we discuss $R$-symmetries in supersymmetry.  In Sec.~\ref{sec:build} we show how to construct a low energy supersymmetric model with an extended $R$-symmetry. In Sec.~\ref{sec:flavor}, we  consider flavor violating observables and the impact of an $R$-symmetry on them.  In particular, we consider the impact of Dirac gauginos on $\Delta F = 2$ processes  in Sec.~\ref{sec:dirac}, on $\Delta F = 1$ processes in Sec.~\ref{sec:fone}, and the impact of the modified Higgs sector in Sec.~\ref{sec:modifiedhiggs}.  CP violation beyond the flavor sector is the topic of Sec.~\ref{sec:CPbeyondflavor}. The effects of small $R$-symmetry violation are considered in Sec.~\ref{sec:approx}. In Sec.~\ref{sec:models}, we discuss specific UV realizations of this scenario, including addressing certain questions of naturalness in these models. In Sec.~\ref{sec:pheno}, we briefly outline the unusual collider phenomenology of these models. Finally, in Sec.~\ref{sec:conc}, we conclude.

\section{$R$-symmetry in Supersymmetry}
\label{sec:rsym}

The supersymmetry algebra automatically contains a continuous $R$-symmetry. 
It was argued long ago \cite{Nelson:1993nf} that the existence of an $R$-symmetry 
in the hidden sector is a necessary condition for supersymmetry breaking. 
A variety of supersymmetric theories exhibit 
supersymmetry breaking without breaking the $R$-symmetry,   notably  the recently discovered 
nonsupersymmetric metastable vacua in supersymmetric gauge theories \cite{Intriligator:2007py,Intriligator:2007cp,Intriligator:2006dd}. Why, then, has unbroken $R$-symmetry not played a larger role in supersymmetric model building?

There are three basic reasons.  The phenomenological lore has
been that gaugino masses require $R$-symmetry breaking.  This is
true for \emph{Majorana} gaugino masses, but 
perfectly viable \emph{Dirac}
gaugino masses (see \cite{Polchinski:1982an, Dine:1992yw, Fox:2002bu}) are possible when the gaugino is paired up
with the fermion from a chiral superfield in the 
adjoint representation. Similarly, the $\mu$ term also breaks $R$-symmetry,
in the presence of the $B_\mu$ term, and is also needed to give the Higgsinos a mass.

The second reason is that models of dynamical supersymmetry breaking
generally break the $R$-symmetry.   However, as already alluded to above, nonsupersymmetric vacua do not always break the $R$-symmetry.  For example, O'Raifeartaigh models may preserve an $R$-symmetry,
and, intriguingly, some simple models of supersymmetry breaking in  
meta-stable vacua also preserve the $R$-symmetry, for  a review see \cite{Intriligator:2007cp}.

The last reason is related to embedding supersymmetry breaking in
supergravity.  At the very least, two conditions must be satisfied:
the gravitino must acquire a mass, and the cosmological constant 
must be tunable to (virtually) zero.  The second condition is 
usually satisfied by adding a constant term in the superpotential,
breaking the $R$-symmetry explicitly.  Indeed, it is this term
that ensures the $R$-axion that results from a spontaneously broken 
$R$-symmetry is given a small but non-zero mass \cite{Bagger:1994hh}.
There are potential loopholes to this generic argument, however. 
One is that, in some cases, the cosmological constant could also be 
canceled by fields in the K\"ahler 
potential that acquire large expectation values \cite{Claudson:1983cr}. 
Second,
we show in Sec.~\ref{sec:approx} that even with only an approximate $R$-symmetry, with small $R$-violating effects 
(as in  the ``supersymmetry without supergravity" framework of \cite{Luty:2002ff}, \cite{Goh:2003yr}), much of the benefits to reducing the supersymmetric contributions
to flavor violation carry through.

\section{Building an $R$-symmetric supersymmetric model}
\label{sec:build}

Our starting point is thus supersymmetry breaking originating 
from hidden sector spurions that preserve the $R$-symmetry.
Both $F$-type and $D$-type supersymmetry breaking is allowed, 
which we can write in terms of the spurions 
$X = \theta^2 F$ and $W'_\alpha = \theta_\alpha D$,
where the $R$-charge assignments of the spurions are necessarily $+2$ and $+1$ 
respectively.  The $W'$ can be considered a hidden sector
$U(1)'$ that acquires a $D$-term.  We assume that the sizes
of the $F$-type and $D$-type breaking are roughly comparable
up to an order of magnitude or so.
Coupling these spurions in an $R$-preserving manner to a low energy supersymmetric 
theory  gives rise to the most general theory with softly broken supersymmetry and an  $R$-symmetry.

Assuming ordinary Yukawa couplings are $R$-symmetric,
and that electroweak symmetry breaking expectation values  
$\langle H_{u,d} \rangle$   do 
not break $R$-symmetry,
the quark and lepton superfields must have $R$-charge $+1$
and the Higgs superfields have $R$-charge $0$.
Gauge superfields $W_i$ have their usual $R$-charge $+1$.

For the MSSM, writing all operators consistent with the 
SM gauge symmetries and the extended $R$-symmetry, we find:
\begin{itemize}
\item Majorana gaugino masses are forbidden.
\item The $\mu$-term, and hence Higgsino mass, is forbidden.
\item $A$-terms are forbidden.
\item Left-right squark and slepton mass mixing is absent
(no $\mu$-term and no $A$-terms).
\item The dangerous $\Delta B = 1$ and $\Delta L = 1$ 
operators, $Q_L L_L D_R$, $U_R U_R D_R$, $L_L L_L E_R$, and $H_u L_L$, are forbidden.
\item Proton decay through dimension-five operators,
$Q_L Q_L Q_L L_L$ and $U_R U_R D_R E_R$, is forbidden
\footnote{Dimension-five proton decay operators are forbidden
by a continuous $R$-symmetry as well as $Z_{2n}$ with $2n \ge 6$
but not by $Z_4$, see e.g. \cite{Kurosawa:2001iq}}.
\item $\Delta L = 2$ Majorana neutrino mass, $H_u H_u L_L L_L$, is allowed.
\end{itemize}

Already we see that the extended $R$-symmetry leads to several improvements 
over the MSSM.  However, the MSSM gauginos and Higgsinos 
are massless, in obvious conflict with experiment.  We must therefore 
augment the MSSM in such a way that allows for $R$-symmetric gaugino 
and Higgsino masses.

\subsection{Gaugino masses}
 
The first obstacle to overcome is to generate a gaugino mass.  Remarkably,
$R$-symmetric gaugino masses are possible when the gauginos are Dirac.
Such a possibility has been explored in a number of contexts previously. For instance, in \cite{Hall:1990hq,Randall:1992cq}, gluinos were made Dirac by adding a color octet, and electroweak gauginos acquired their masses via marrying the superpartners of the Goldstone modes in the Higgs supermultiplet. In \cite{Fox:2002bu}, Dirac gauginos were motivated as an ultraviolet insensitive and flavor blind means of mediating SUSY breaking, which resulted in the so-called ``supersoft'' spectrum with gauginos a factor of $(4\pi/\alpha)^{1/2}$ above the scalars. They have additionally been considered in a variety of phenomenological contexts recently \cite{Nelson:2002ca,Chacko:2004mi,Carone:2005iq,Nomura:2005rj,Nomura:2005qg,Nakayama:2007cf,Antoniadis:2006uj,Hsieh:2007wq}.

Unlike previous attempts to implement Dirac gauginos within the context of flavor-blind SUSY breaking masses, we will simply consider them an element of a general softly broken supersymmetric theory, which may also contain soft masses from other sources for the scalars.  Dirac gauginos require the addition of an adjoint chiral superfield $\Phi_i$ to the theory for each gauge group $i$. Then the $R$-symmetric operator involving a $D$-type spurion is \cite{Dine:1992yw,Fox:2002bu}:
\be
\label{diracmassoperator}
  \int d^2 \theta\; \frac{W_\alpha'}{M}\; W^\alpha_i \Phi_i~,
\ee
which leads to a Dirac mass for each gaugino $m_i \lambda_i \psi_i$.  Here $i = \tilde{B}, \tilde{W}, \tilde{g}$ and $m_i \propto D/M$, pairing up the two-component gaugino with the two-component fermion in the chiral adjoint.  The mediation scale, $M$, of supersymmetry breaking from the hidden sector to the visible sector, could be as high as $\MPl$ (as in gravity-mediation), or a much lower scale (as we discuss in Sec.~\ref{sec:UVcompletion}).

\subsection{Extended Higgs sector}

The second obstacle is the absence of a $\mu$-term.  
Aside from the approach of  \cite{Nelson:2002ca}, in which the Higgsinos acquired a mass without an explicit $\mu$ term, 
the only option is to enlarge the Higgs sector. 
This can be done by adding multiplets $R_u$ and $R_d$ 
that transform under $SU(2)_L \times U(1)_Y$ the same way as 
$H_d$ and $H_u$, respectively, except that they have $R$-charge $+2$.
This allows the following supersymmetric mass terms:
\be
\label{Wmu}
W_\mu = \mu_u H_u R_u + \mu_d H_d R_d~.
\ee
These mass terms can be thought of as arising naturally from the
Giudice-Masiero mechanism:
\be
\int d^4\theta \frac{X^\dag}{M} H_u R_u + \frac{X^\dag}{M} H_d R_d~,
\ee
automatically explaining why their size is near to the scale of 
soft supersymmetry breaking.  Also, the scalar components of the 
Higgs acquire expectation values that break electroweak symmetry, 
while the $R$'s do not, thus preserving the $R$-symmetry.

A holomorphic $B_\mu$-term is consistent with the $R$-symmetry 
assignments.  It too can be naturally generated of the right
size through the ordinary Giudice-Masiero mechanism,
\be
\int d^4\theta \frac{X^\dag X}{M^2} H_u H_d \; .
\ee

\subsection{Soft masses and other interactions}

Nonholomorphic soft terms---scalar masses for the squarks, 
sleptons, Higgses, Higgs partners $R_u$,$R_d$, and 
$\Phi_{\tilde{B},\tilde{W},\tilde{g}}$---are allowed through the usual operators:
\be
\int d^4 \theta \frac{X^\dag X}{M^2} Q_i^\dag Q_j + \ldots + 
\frac{X^\dag X}{M^2} H_u^\dag H_u + \ldots 
\ee
For brevity we have only written the soft terms for $Q$ and $H_u$
while the other terms are analogous.  Note that flavor-violation
may be arbitrarily large in the squark and slepton sector
since we assume no particular flavor structure of these operators.
Also, nonholomorphic scalar mass mixing from operators 
such as $X^\dag X H_u^\dag R_d$ is forbidden by the $R$-symmetry.

Other holomorphic soft terms consistent with the $R$-symmetry 
assignments include masses for the scalar components of the 
three adjoints:
\be
\label{adjsoft}
\int d^4\theta  \frac{X^\dag X}{M^2} \, {\rm tr} \, \Phi_i^2+
\int d^2\theta   \frac{W'^\beta W_\beta'}{M^2} \, {\rm tr} \, \Phi_i^2 \; .
\ee

Finally, there is another set of supersymmetric couplings allowed 
by the $R$-symmetry---the couplings of the $\Phi_i$-adjoint chiral 
superfields to the Higgs doublets,
\be
\label{Wphi}
W_\Phi =\sum\limits_{i=\tilde{B},\tilde{W}} \lambda_u^i H_u \Phi_i  R_u 
+ \lambda_d^i R_d \Phi_i H_d~,
\ee
where $i=\tilde{B},\tilde{W}$ refer to the couplings of the $U(1)_Y$ or $SU(2)_L$ adjoints, respectively \footnote{If the Higgses come from the adjoint of $SU(3)^3$, as we discuss in section \ref{sec:models}, then such terms may be small due to the $N=2$ symmetry of the extended gauge sector.}.

\begin{table}
\begin{tabular}{c|cc}
Field              & $(SU(3)_c,SU(2)_L)_{U(1)_Y}$   & $U(1)_R$ \\ \hline
$Q_L$              & $({\bf 3},{\bf 2})_{1/6}$      & 1 \\ 
$U_R$              & $({\bf \bar 3},{\bf 1})_{-2/3}$     & 1 \\ 
$D_R$              & $({\bf \bar 3},{\bf 1})_{1/3}$      & 1 \\ 
$L_L$              & $({\bf 1},{\bf 2})_{-1/2}$     & 1 \\ 
$E_R$              & $({\bf 1},{\bf 1})_{1}$        & 1 \\ 
$\Phi_{\tilde{B}}$ & $({\bf 1},{\bf 1})_{0}$        & 0 \\ 
$\Phi_{\tilde{W}}$ & $({\bf 1},{\bf 3})_{0}$        & 0 \\ 
$\Phi_{\tilde{g}}$ & $({\bf 8},{\bf 1})_{0}$        & 0 \\ 
$H_u$              & $({\bf 1},{\bf 2})_{1/2}$      & 0 \\ 
$H_d$              & $({\bf 1},{\bf 2})_{-1/2}$     & 0 \\ 
$R_u$              & $({\bf 1},{\bf 2})_{-1/2}$     & 2 \\ 
$R_d$              & $({\bf 1},{\bf 2})_{+1/2}$     & 2 
\end{tabular}
\caption{Matter and $R$-charges in the $R$-symmetric supersymmetric
model.}
\label{table:matter}
\end{table}

Together these three elements (Dirac gauginos, zero $A$-terms, and the modified Higgs sector) allow an $R$-symmetric theory to be written. We will refer to this theory, with an extended $R$-symmetry, as the Minimal $R$-symmetric Supersymmetric Model (MRSSM). The matter superfields and $R$-symmetry assignments are given in Table \ref{table:matter}.  We note that the $R$ symmetry of the resulting model is free of gauge anomalies.  Interestingly, even if we enforce only a partial $R$-symmetry on low energy supersymmetry, many benefits for flavor violating signals persist. One could include only the Dirac gaugino masses with a standard $\mu$ term (for instance as in \cite{Fox:2002bu,Carpenter:2005tz,Chacko:2004mi,blechinprog}) or just the extended Higgs sector. We shall see that the Dirac gauginos together with no $A$-terms tend to address flavor problems at small $\tan \beta$ while the extended Higgs sector addresses flavor problems at large $\tan \beta$.

\bigskip

\section{Flavor with an Extended $R$-symmetry}
\label{sec:flavor}

There are many different searches for flavor violation in precision observables, with many different sources in supersymmetric theories. There are $\Delta F=2$ processes, such as contributions to meson mass differences from mixing (i.e., $K$--$\bar{K}$ and $B$--$\bar{B}$ mixing), as well as $\Delta F=1$ processes, such as $b \rightarrow s \gamma$ or $\mu \rightarrow e \gamma$. In supersymmetric theories, these can arise from a number of diagrams, including diagrams involving gauginos, radiative corrections to Higgs couplings, or charged Higgs bosons. In this section, we will attempt to separate the flavor-violating effects of Dirac gauginos from those of the absence of $A$-terms and of a modified Higgs sector. 

In general, we find that the presence of Dirac gauginos and absence of $A$-terms ameliorate problems of flavor over a wide range in $\tan \beta$, and are both essential for any value of $\tan \beta$ if $O(1)$ flavor violation is to be allowed. At large $\tan \beta$, there are additional diagrams in the MSSM \cite{Hamzaoui:1998nu} which become important to FCNCs. These diagrams are eliminated by extending the Higgs sector to one with $R$-symmetric $\mu$ terms, thus altogether allowing $O(1)$ flavor violation over the entire range in $\tan \beta$.

\subsection{Flavor Violation with Dirac Gauginos}
\label{sec:dirac}

Any process in the MSSM which involves gauginos propagating in the loop can be affected by the presence of Dirac, as opposed to Majorana, gauginos. We can loosely separate those into $\Delta F=2$ and $\Delta F=1$ pieces.

\begin{figure*}
\begin{center}
\includegraphics[width=3.2in]{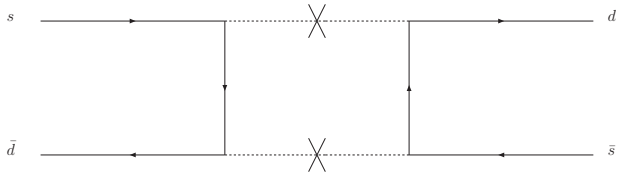} 
 \end{center}
 \caption{Box diagram contributing to $K-\bar{K}$ mixing.}
\label{fig:box}
\end{figure*}

\subsubsection{$\Delta F =2$ Flavor Violation}

The most stringent constraints on flavor violation come from studies of the kaon system. That the observed $K_L$-$K_S$ mass difference is well explained by standard model physics places severe constraints on flavor violation in the squark soft mass  squared matrices. In the MSSM, diagrams such as 
Fig.~\ref{fig:box} with $O(1)$ flavor violation contribute well in excess of the experimental limits. 
Consider first the contribution to flavor violation from gluinos.
For  $s$-$d$ flavor violation, if the flavor violation is only in the right- or left-handed squarks, the limits are \cite{Ciuchini:1998ix}:
\be
\delta_{LL}, \delta_{RR} \; \lsim \; 4.6 \times 10^{-2} .
\ee
In the presence of both left- and right-handed flavor violation, the limits are more severe:
\be
\sqrt{\delta_{LL} \delta_{RR}} \; \lsim \; 9.6 \times 10^{-4} .
\ee
All results are quoted for $m_{\tilde g} = m_{\tilde q} = 500$~GeV.

\begin{figure*}
a)\includegraphics[width=2.2in]{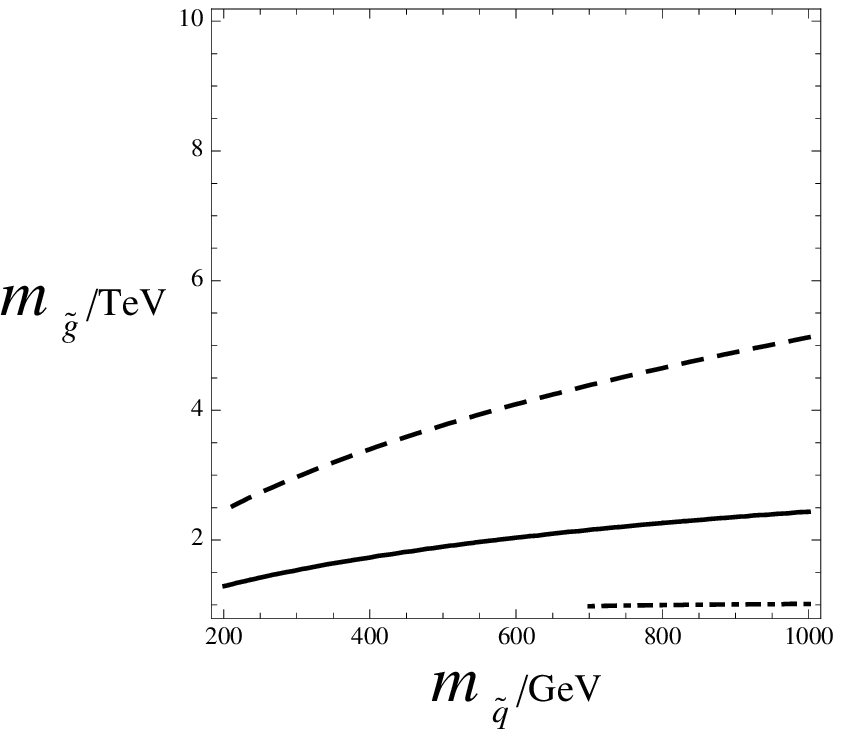} \hskip 1in
b)\includegraphics[width=2.2in]{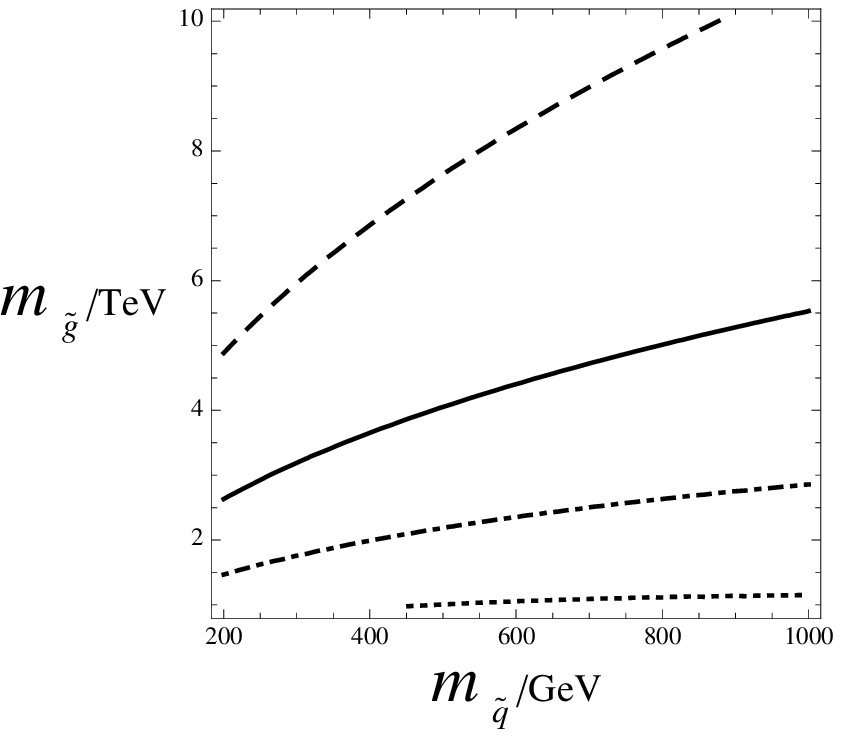}
\caption{Contours of $\delta$ where $\Delta M_{\rm box} = \Delta M_k$ for a) $\delta_{LL}=\delta$, $\delta_{RR}=0$, b) $\delta_{LL}=\delta_{RR}=\delta$.  An identical plot to a) exists for $\delta_{LL}=0$, $\delta_{RR} =\delta$. Contours are $\delta =0.03,0.1,0.3,1$ for dotted, dot-dashed, solid, dashed respectively.}
\label{fig:kkbarlimits} 
\end{figure*}

In the $R$-symmetric model, the contributions to flavor-violating processes are significantly reduced due to two main effects.  First, the radiative corrections to squark masses from Dirac gauginos are finite one-loop effects, unlike Majorana gauginos that lead to a one-loop log-enhanced effects familiar in the MSSM.  Dirac gauginos can therefore be naturally heavier than squarks by about a factor of about $10$.  This increase in the gaugino mass implies that flavor-violating observables are suppressed by $m_{\tilde q}^2/m_{\tilde g}^2 \sim 10^{-2}$ in an $R$-symmetric model, as compared with squarks and gluinos that are inevitably similar in mass in the MSSM.
 
If that alone were sufficient to suppress the box diagram, it would have been considered, even with unnatural tuning, in $R$-violating supersymmetry. However, the presence of the $R$-symmetry goes further. Ordinarily, integrating out the Majorana gluinos gives dimension five operators such as:
\be
\label{dim5gaugino}
\frac{1}{m_{\tilde g}}\; \tilde d_R^* \tilde s_L^* \bar d_R s_L.
\ee
The $R$-symmetry forbids these dimension five operators, and the leading operators are dimension six, such as:
\be
\frac{1}{m_{\tilde g}^2}\;  \tilde d_L \partial_\mu \tilde s^*_L \; \bar d_L \gamma^\mu s_L.
\ee
The box diagrams are dominated by momenta $k_{\rm box} \sim m_{\tilde q}$, which leads to an additional overall suppression of $m_{\tilde q}^2/m_{\tilde g}^2 \sim 10^{-2}$. Together, these effects lead to a sizeable suppression of the box diagram, allowing order one flavor violating soft masses, even for relatively light squarks.

In the presence of Dirac gauginos, the box diagram yields a contribution the the $K$--$\bar{K}$ mass difference:
\be
\Delta M_{\rm box} = 2 (C_1 M_1 + C_4 M_4 + C_5 M_5)~,
\ee
where:
\bea
C_1 &=& \frac{\alpha_s^2}{216 m_{\tilde q}^2} (\delta_{LL}^2 + \delta_{RR}^2) 66 \tilde f_6(x)~, \nonumber \\
C_4 &=& -\frac{\alpha_s^2}{216 m_{\tilde q}^2} (72 \delta_{LL}  \delta_{RR}) \tilde f_6(x)~, \\
C_5 &=& \frac{\alpha_s^2}{216 m_{\tilde q}^2} (120 \delta_{LL} \delta_{RR}) \tilde f_6(x),\nonumber 
\eea
\be
\label{tildef6}
\tilde f_6(x) = \frac{6x(1+x) \log(x) - x^3 - 9x^2+9x +1}{3(x-1)^5}, \nonumber
\ee
and
\bea
\label{m145}
M_1 &=& \frac{1}{3} m_K f_K^2 B_1~, \nonumber \\
M_4 &=& \left( \frac{1}{24} + \frac{1}{4}\left(\frac{m_K}{m_s+m_d} \right)^2 \right) m_K f_K^2 B_4~, \\
M_5 &=& \left( \frac{1}{8} + \frac{1}{12}\left(\frac{m_K}{m_s+m_d} \right)^2 \right) m_K f_K^2 B_5~. \nonumber
\eea
Here $x = m_{\tilde g}^2/m_{\tilde q}^2$, $B_{1,4,5} = 0.6,1.03,0.73$ are bag factors for the relevant operators. Our numbering is chosen to be consistent with \cite{Bagger:1997gg} (the coefficients $C_{2,3}$ for operators $O_{2,3}$, in their numbering, vanish in the absence of left-right mixing).

In Fig.~\ref{fig:kkbarlimits}, we show the constraints on the $\delta$'s by requiring that the new physics contributions are smaller than the observed value \footnote{It is important to note that we have not calculated leading QCD corrections to these contributions as in \cite{Bagger:1997gg}. However, because the relevant operators here are the new dimension six operators and not the dimension five usually present in SUSY, the corrections found there can not be simply applied here.}.  Immediately one can see a remarkable change from usual SUSY theories. First of all, $O(1)$ flavor violation is allowed for few TeV mass Dirac gauginos, where it would be completely excluded for similar mass Majorana gauginos in the MSSM. Second, the limits on flavor violation weaken as the squark mass is decreased, whereas they would generally strengthen in the MSSM.

Although we expect Winos and Binos to be significantly lighter than gluinos, their presence in the loops should not radically change our results. Wino box diagrams only contribute to the limits on $\delta_{LL}^2$ which are much weaker than $\delta_{LL} \delta_{RR}$. Bino box contributions to $\delta_{LL} \delta_{RR}$ terms have a suppression of $(1/3)^2 \times (1/6)^2 g_Y^4$, which, even neglecting additional color factor enhancements, would require Binos approximately ten times lighter than gluinos in order for the contributions to be competitive. While it would be interesting to determine the precise bounds on the Bino mass, the effects are likely to be less significant than the leading QCD corrections which are not included here.

This setup is a radical departure from previous approaches to the flavor problem.  The severe limits in the MSSM required that either a flavor-blind mediation mechanism was at work, enforcing all off-diagonal elements to be extremely small, or to otherwise raise the masses of the offending squarks to extremely high values, as in effective supersymmetry.  Here, a combination of the natural ability to raise the gluino mass above the squark mass, combined with an additional suppression in the box diagrams coming from the Dirac nature of the gluinos, allows one to consider genuinely large and experimentally accessible flavor violation, even in the first two generations.

\begin{figure*}
a)\includegraphics[width=2.2in]{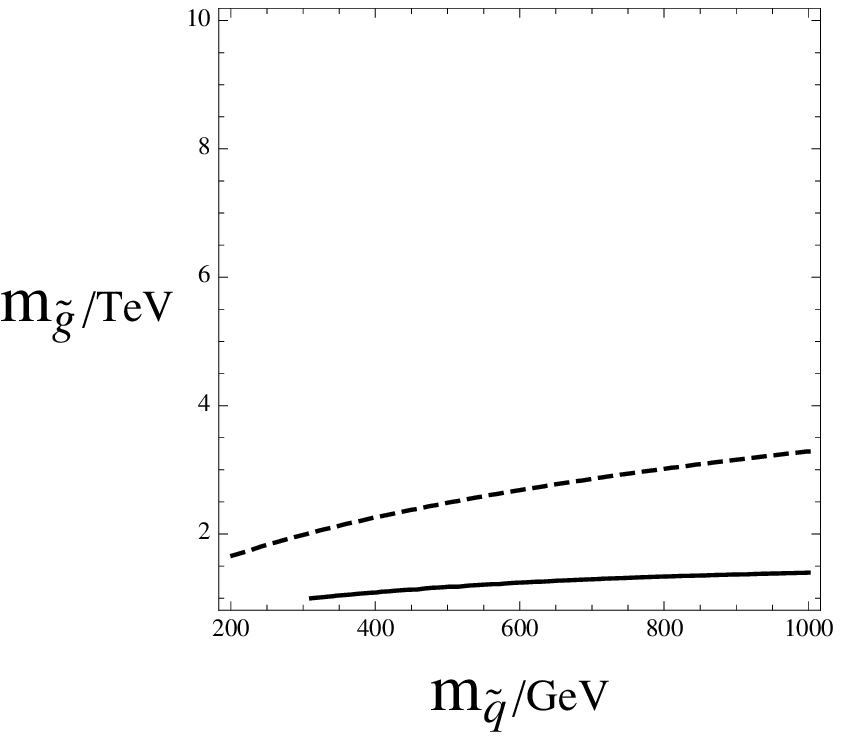} \hskip 1in
b)\includegraphics[width=2.2in]{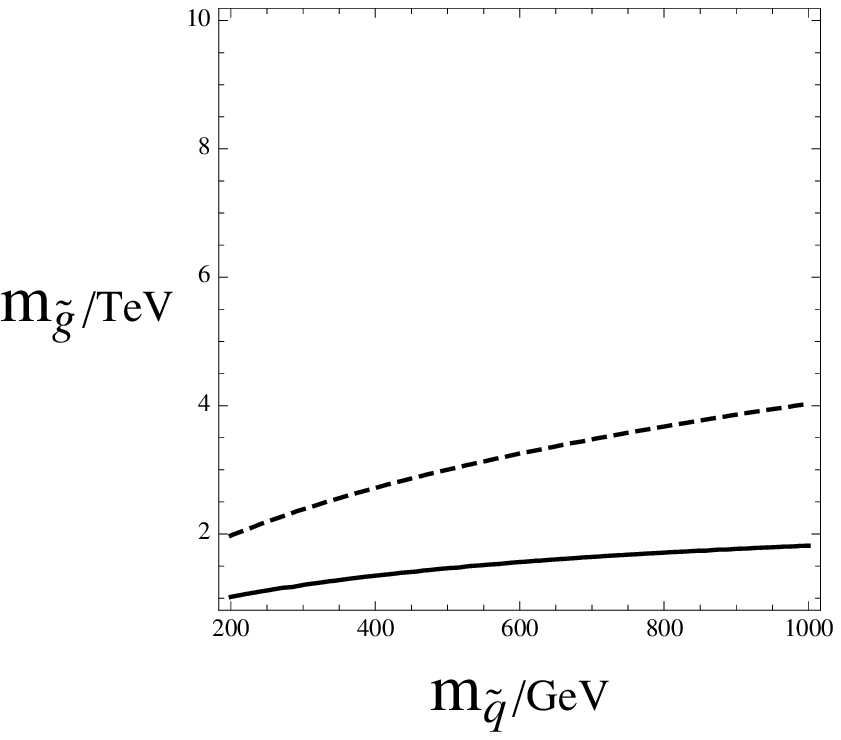}
\caption{Same as Fig.~\ref{fig:kkbarlimits} but for $B_d$ mesons. Contours of $\delta$ where $\Delta M_{\rm box} = \Delta M_{B_d}$ for a) $\delta_{LL}=\delta$, $\delta_{RR}=0$, b) $\delta_{LL}=\delta_{RR}=\delta$. An identical plot to a) exists for $\delta_{LL}=0$, $\delta_{RR} =\delta$. Contours are $\delta =0.1,0.3,1$ for dot-dashed, solid, dashed respectively.}
\label{fig:bmix} 
\end{figure*}

\subsubsection{$\epsilon_K$}

Even stronger constraints exist on the imaginary parts of the flavor violation in supersymmetry. In particular, $\epsilon_K$, with a measured value of $2.229 \times 10^{-3}$ \cite{Yao:2006px} limits the imaginary component of the operators considered above to be smaller by $6.3 \times 10^{-3}$ than the flavor conserving pieces (taking the simplified limit in which the beyond the standard model contribution saturates the observed value). In our scenario, there are two basic approaches to $\epsilon_K$: one can invoke a moderate flavor degeneracy (which can be natural in some regions of parameter space), or one can insist that the flavor violating soft masses are real.

If we consider imaginary squark masses, we must isolate the physical phases. In the squark sector of the MRSSM, physical phases exist in the Yukawa couplings and squark mass matrices.  It is straightforward to count them:  Each quark Yukawa matrix ($Y_U,Y_D$) has 9 complex phases, giving a total of 18 new potential phases.  Performing global U(3)$^3$ rotations on the quark superfields removes all but one physical phase corresponding to the unbroken global symmetry U(1)$_B$. This leaves one phase in the CKM matrix and all phases in the squark (mass)$^2$ matrices being physical.  

If we allow imaginary contributions to the flavor violating mass terms, $O(1)$ off diagonal corrections to the soft masses would not be allowed if phases are also large. However, if there is a moderate flavor degeneracy, then reasonable phases are allowed. One can read the strongest constraints from Fig.~\ref{fig:kkbarlimits}(b), simply by reading the contours as more stringent by a factor of $6.3 \times 10^{-3}$ on ${\rm Im}(\delta_{LL} \delta_{RR})$. (More precisely, a contour of $\delta = 1$ can be thought of as a contour of ${\rm Im}(\delta_{LL} \delta_{RR}) = 1^2 \times( 6.3 \times 10^{-3})$, a contour of  $\delta = 0.3$ can be thought of as a contour of ${\rm Im}(\delta_{LL} \delta_{RR}) = 0.3^2 \times (6.3 \times 10^{-3})$, etc.)
For example, consider a gluino mass of $3.5$~TeV and a squark mass of $400$~GeV. With off diagonal elements of size $100$~GeV ($200$~GeV), corresponding to $\delta = 0.06\; (0.25)$, the phase is bounded to be $\theta <0.15\; (0.01)$.

Such a moderate suppression of off-diagonal contributions to squark masses can be natural in the MRSSM in certain regions of parameter space, given the finite one-loop flavor-blind contributions from Dirac gauginos.  For instance, consider that flavor-arbitrary soft masses for all sfermions are of order $100-200$~GeV. Squarks receive a finite contribution of $O(m_{\tilde{g}}/5)$ from the Dirac gluinos, while corrections to sleptons from Winos/Binos are much smaller, leaving larger relative flavor violation there. This would render the above example completely natural in the MRSSM.

A second approach is to assume that the soft scalar masses, although flavor violating, are real. This could arise if CP is a symmetry of the SUSY breaking sector, for instance. However, it is conceivable that the soft scalar masses squared might all be real (i.e., have no relative phase) because the operators are all of the form $X^\dagger X Q^\dagger Q$, even absent CP in the SUSY breaking sector. In contrast, while the operators generating $\mu$, $B_\mu$ and Dirac masses all have dramatically different forms, making their phases unlikely to be equal, absent some symmetry reason.

\subsubsection{B meson mixing}

Just as box diagram contributions to $K$--$\bar{K}$ mixing are suppressed, so, too, are they suppressed for B meson mixing. The above calculations carry over to the $B$ meson case trivially if one replaces the appropriate quark and meson masses, and bag factors.

We find that for the parameters listed above, the contributions should be much smaller than the recently measured $\Delta M_{Bs} = 17.77 \pm 0.10 (stat) \pm 0.07 (syst) {\rm ps}^{-1}$ \cite{Abulencia:2006mq,Abazov:2006dm,Abulencia:2006ze}. However, there is the possibility of significant contributions to $B_d$ mixing, which constrain the flavor violation, although more weakly than that of $K-\bar K$ mixing. In Fig.~\ref{fig:bmix}, we show the equivalent plot of Fig.~\ref{fig:kkbarlimits}, but for the case of $B_d$ oscillations.  Clearly, no significant constraint from $B_d$ mixing on the relevant flavor-violating $\delta$s is present in the MRSSM.

\subsection{$\Delta F=1$ Flavor Violation}
\label{sec:fone}

Flavor changing processes such as $\mu \rightarrow e \gamma$ or $b \rightarrow s \gamma$ involve a helicity flip in the diagram. For Dirac gauginos, the opposite helicity state has no direct couplings to sfermions, so the diagram with a helicity flip on the gaugino line is absent \cite{Fox:2002bu,Nelson:2002ca}. This leaves only the much smaller diagram where the helicity flip occurs on the external quark or lepton line, or a helicity flip on the internal line coming from Higgsino-gaugino mixing. As a consequence, we shall see that large flavor violation is allowed for these $\Delta F=1$ processes as well.

\subsubsection{$\mu \rightarrow e \gamma$}

\begin{figure}
\begin{center}
\includegraphics[width=2.2in]{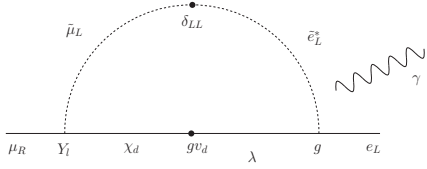}
\end{center}
\caption{Internal (Yukawa) chirality flip diagram contributing to $\mu \rightarrow e \gamma$; $\chi_d$, $\lambda$ denote the appropriate Dirac Higgsino and gaugino and the photon here and in Fig.~\ref{fig:LFVexternal} can be attached to any charged line. }
\label{fig:LFVinternal}
\end{figure}

\begin{figure}
\begin{center}
\includegraphics[width=2.2in]{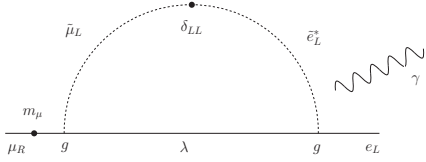}
\end{center}
\caption{External chirality flip diagram contributing to $\mu \rightarrow e\gamma$;  $\lambda$ denotes  the appropriate Dirac gaugino.}
\label{fig:LFVexternal}
\end{figure}

\begin{figure*}
a)\includegraphics[width=2.2in]{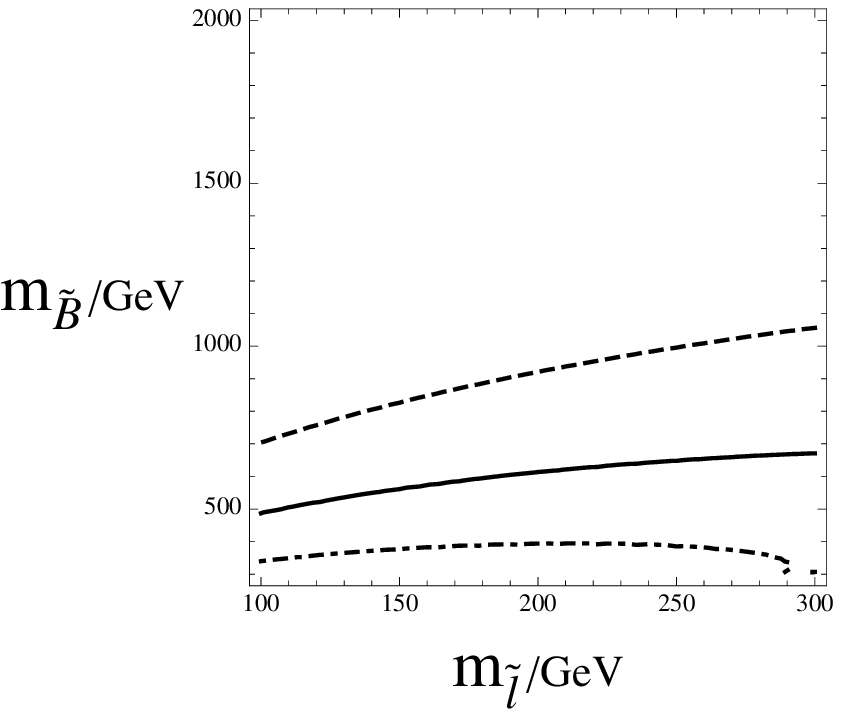}\hskip 1in
b)\includegraphics[width=2.2in]{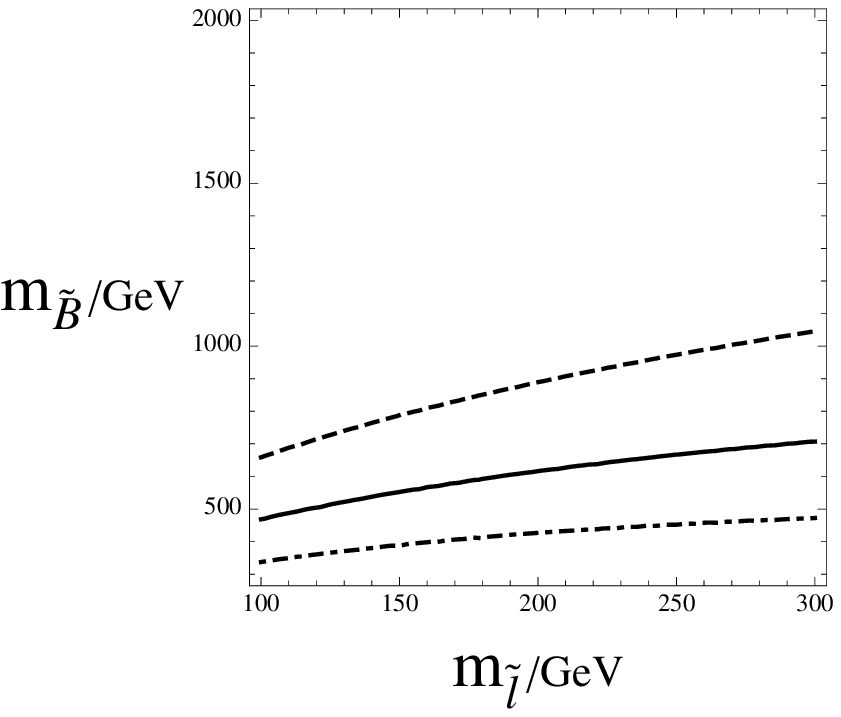}
\caption{Contours of $\delta$ where $BR_{\mu \rightarrow e \gamma} =1.2 \times 10^{-11}$ for a) $\delta_{LL}=\delta$, $\delta_{RR}=0$, b) $\delta_{RR}=\delta$, $\delta_{LL}=0$. Contours are $\delta =0.1,0.3,1$ for dot-dashed, solid, dashed respectively, for $m_{\tilde{B}} = m_{\tilde{W}}/2$. }
\label{fig:muegamma} 
\end{figure*}

The most stringent constraint on $\Delta F=1$ flavor changing is from $\mu \rightarrow e \gamma$. There are two types of diagrams contributing to lepton flavor violation in the $R$-symmetric model, shown on figs.~\ref{fig:LFVinternal} and \ref{fig:LFVexternal}. The $\mu \rightarrow e \gamma$ branching ratio is given by \cite{Hisano:1998fj}:
\bea
 BR_{\mu \rightarrow e \gamma} &=& \frac{48 \alpha \pi^3}{G_F^2} \left(|A_{L
c1} + A_{L n1} + A_{L c2} + A_{L n2}|^2\right. \nonumber \\\
 && \left. \; +\;  |A_{R n1} + A_{R n2}|^2 \right)~,
\eea
where the amplitudes due to  graphs with chargino (neutralino) exchange and chirality flip in the  external line are denoted by a subscript $c1 (n1)$:
\beq
\label{lfvexternal}
A_{L c1} &=& \frac{\alpha_2}{24 \pi} \frac{\delta_{LL}}{m^2_{\tilde{l}} }\;
G_{c1}(x_c)~,\nonumber \\
A_{L n1} &=& - \frac{\alpha_2}{48 \pi} \frac{\delta_{LL}}{m^2_{\tilde{l}} }\;
(1 +   \tan^2 \theta_W) \; G_{n1}(x_n)~,\nonumber \\
A_{R n1} &=& - \frac{\alpha_1}{12 \pi} \frac{\delta_{RR}}{m^2_{\tilde{l}} } \;
G_{n1}(x_n)~,
\eeq
with $x_c = m_{c}^2/m^2_{\tilde{l}}$, $x_n = m_{n}^2/m^2_{\tilde{l}}$ where $m_{c}$ and $m_{n}$ are the chargino and Dirac neutralino mass eigenstates.  Similarly, the contributions of graphs with an internal chirality flip are denoted by a subscript $c2(n2)$ and are as follows:
\beq
\label{lfvinternal}
A_{L c2} &=&   - \frac{\alpha_2 }{4 \pi }
\frac{\delta_{LL}}{m^2_{\tilde{l}} } \;   G_{c2}(x_c) \nonumber \\
A_{L n2} &= &  \frac{\alpha_2 }{16 \pi }
\frac{\delta_{LL}}{m^2_{\tilde{l}} }   (1-   \tan^2 \theta_W)
G_{n2}(x_n)    \nonumber \\
A_{R n2} &=& \frac{\alpha_1}{8 \pi }
\frac{\delta_{RR}}{m^2_{\tilde{l}} }  G_{n2}(x_n)~.
\eeq
Here all mixing angles were calculated to leading order in an expansion in $m_W/m_{\tilde{B}}$, $m_W/m_{\tilde{W}}$ and the various functions $G$ are given by:
\bea
\label{gfunctions}
G_{n1}(x) 
&=&  \frac{17 x^3-9x^2-9x+1-6x^2(x+3) \log x }{2(1-x)^5}~,   \nonumber \\
G_{c1}(x) &=& \frac{x^3+9 x^2-6x (x+1) \log x   -9 x-1}{(x-1)^5}~, \nonumber \\
G_{n2}(x)&=&\frac{1+4x-5x^2+4x\log x +2 x^2 \log x}{ (1-x)^4} ~,\nonumber \\
G_{c2}(x)&=& \frac{x^2+4 x-2 (2 x+1) \log x -5}{2 (x-1)^4} ~.
\eeq
On Fig.~\ref{fig:muegamma}, we show the limits on chargino/neutralino and squark masses from $\mu\rightarrow e\gamma$. 

Interestingly, the most significant diagrams at large $\tan \beta$ in the MSSM involve both $\mu$ and Majorana gaugino insertions (see Fig.~12 of \cite{Hisano:1998fj}). Thus, the relative weakness of the constraint of $\mu \rightarrow e \gamma$ in this framework is a combination of the heavier gauginos, lack of $A$-terms, and then either of the Dirac nature of the gauginos or the modified Higgs sector.

\subsubsection{$b\rightarrow s\gamma$}

The contribution to $b \rightarrow s \gamma$ branching ratio is \cite{Gabbiani:1988rb,Gabbiani:1996hi}:
\bea
BR_{b\rightarrow s \gamma}=
\frac{\alpha_s^2 \alpha m_b^3  \tau_B}{81 \pi^2 m_{\tilde q}^4}  \,
\left( \left |  m_b \frac{G_{n1}(x)}{6} \delta_{LL} \right |^2 + L \leftrightarrow R \right) 
\eea
where $G_{ n1}$ is defined in (\ref{gfunctions}).
This contribution is well below the observed value.
For instance, taking $m_{\tilde g}=1.5$~TeV, $m_{\tilde q}=300$~GeV,  we find:
\bea
\delta BR_{b\rightarrow s \gamma} = 1.5 \times 10^{-8} \delta_{LL}^2~.
\eea
In comparison, the world average for $b\rightarrow s \gamma$ with a photon threshold of $E_\gamma > 1.6$~GeV is \cite{Barberio:2006bi}:
\be
BR_{\bar B \rightarrow X_s \gamma} = (3.55 \pm 0.24 ^{+0.09}_{-0.10}\pm 0.03) \times 10^{-4}.
\label{eq:bsgbound}
\ee
Technically, a more precise bound can be obtained by keeping the interference term with the calculable sizeable standard model contribution, see e.g. 
\cite{Fox:2007in}; however, since this process clearly poses no significant contraints on our framework, we do not consider this in more detail here.  
Note that while we considered only gluino contributions to the decay $b \rightarrow s\gamma$, the contributions from Winos are even smaller (once the bounds from lepton flavor violation are included).  Also, given the smallness of the corrections to rare B decays, significant CP asymmetries from SUSY contributions are highly unlikely.

\subsubsection{$\epsilon'/\epsilon$}

In the MSSM, the CP violating observable $\epsilon'/\epsilon$ also constrains the presence of CP violation in new physics. The strongest constraints are on left-right insertions, with a limit of $|{\rm Im}(\delta_{LR})| < 2 \times 10^{-5}$ for $m_{\tilde g} = m_{\tilde q} = 500$~GeV \cite{Masiero:2002xj}. Left-left insertions, by contrast, have the relatively weak constraint of  $|{\rm Im}(\delta_{LL})| < 4.8 \times 10^{-1}$ for the same parameters. (It should be noted that this is particularly weak due to a cancellation of box and penguin contributions, and for $m_{\tilde g} = 275$~GeV, $1000$~GeV the limits are $|{\rm Im}(\delta_{LL})| < 1.0, 2.6 \times 10^{-1}$, respectively.)

However, it has been shown that for non-degenerate squarks (in particular, for right handed up squarks split from the down squarks), there can be a sizeable $\Delta I = 3/2$ contribution \cite{Kagan:1999iq}.  These contributions are dependent on the particulars of the spectrum and certain assumptions about the phase. Following \cite{Kagan:1999iq} and taking a spectrum $m_{\tilde d_L}^2 = \tilde m^2$,$m_{\tilde d_R}^2 =\tilde  m^2 (1-\delta)$, and $m_{\tilde u_R}^2 = \tilde m^2 (1+\delta)$, one finds a contribution (with $x = (m_{\tilde g}/m_{\tilde q})^2$, as usual):
\be
\Delta \left( \frac{\epsilon'}{\epsilon} \right) \; = \; - 0.75 \left( \frac{500 \; \gev}{\tilde m} \right)^2 \frac{\delta}{x^2} {\rm Im} (\delta_{LL}) \; .
\ee
Requiring this to be smaller than the observed value of $(1.65 \pm 0.26) \times 10^{-3}$ \cite{Yao:2006px} yields very mild constraints. Taking for illustration ${\rm Im} (\delta_{LL}) = \delta$, we find:
\be
\delta \; \lsim \; 1.2 \times \left(\frac{\tilde m}{500 \; \gev} \right) \left(\frac{x}{25} \right) \; .
\ee

In summary, contributions to $\Delta F=1$ FCNCs are not a strong constraint on SUSY effects, at present, although a global analysis of flavor constraints is clearly warranted \cite{Ciuchini:1997zp}. Nonetheless, there is a charged Higgs in the theory, which can still yield interesting contributions, such as to $b \rightarrow s \gamma$.  Lepton flavor violation, while not at present a strong constraint, may yield an interesting signal as tests improve.

\subsection{Flavor at large $\tan \beta$ with a modified Higgs sector}
\label{sec:modifiedhiggs}

\begin{figure*}
\begin{center}
\includegraphics[width=5.2in]{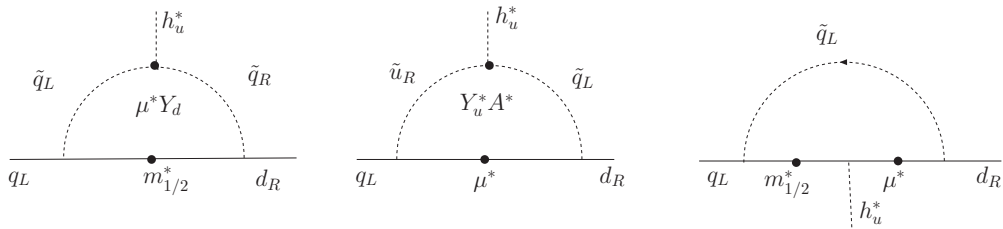}
\end{center}
\caption{The one-loop diagrams contributing to FCNCs at large $\tan\beta$ in the MSSM.  All of these diagrams are absent in the $R$-symmetric model.}
\label{fig:FCNCtanbeta}
\end{figure*}

In the MSSM, couplings of down-type quarks to $H_u$ can be radiatively generated at large $\tan\beta$, giving the largest contribution to FCNCs \cite{Hamzaoui:1998nu}, including mixing effects, but also in decays $B\rightarrow \tau \nu$ or $B_s \rightarrow \mu^+ \mu^-$. The diagrams generating these couplings are shown in Fig.~\ref{fig:FCNCtanbeta}.  As we will now explain, these potentially large contributions are absent in an $R$-symmetric model, with different diagrams eliminated by the absence of $A$-terms, the $\mu$-term, and Majorana gauginos.

To understand the origin of these contributions, recall that the ability to rotate $H_d$ and ($D_R$, $E_R$) with opposite phases corresponds to a $U(1)_{PQ}$ Peccei-Quinn (PQ) symmetry in the MSSM.  If this were exact, the $PQ$ symmetry would forbid the coupling of the up-type Higgs $h_u$ to the down-type quarks.  Alas, $U(1)_{PQ}$ is broken in the MSSM by the superpotential $\mu$-term $\mu H_u H_d$ (as well as the $B_\mu$ term), leading to an important effective dimension-three scalar operator. In the component Lagrangian, this operator couples $h_u^*$ to the down-type squarks:
 \beq
 \label{higgsFCNC1}
 \mu^*  \tilde{q}_L Y_d h_u^* \tilde{d}_R~.
 \eeq 
This interaction violates both the (extended) $R$-symmetry and $PQ$-symmetry, and since it is proportional to the down-type Yukawa coupling, it grows with $\tan\beta$.

The importance of Eq.~(\ref{higgsFCNC1}) at large $\tan\beta$ for flavor-violation is easiest to understand by taking the limit of large gaugino (and possibly Higgsino) masses.  Integrating out a large gluino Majorana mass $m_{\tilde{g}}$ generates tree-level dimension-five operators of the form (\ref{dim5gaugino}): 
\beq
\label{higgsFCNC2}
 { 4 \pi \alpha_3 \over m_{\tilde{g}}} ~\; q_L   d_R \; \tilde{q}_L^* \tilde{d}_R^* ~,
\eeq
where we use 2-component notation for fermions here and in the rest of this section.  
These terms violate the $R$-symmetry but are $PQ$ symmetric (recall that the quark fields have $R$-charge zero, while the squarks have unit $R$-charge, in accordance to our convention from  Sec.~\ref{sec:build}). Combining Eq.~(\ref{higgsFCNC2}) with the $R$ and $PQ$-violating interaction (\ref{higgsFCNC1}), and closing the squark lines into a loop, we obtain a coupling of the form 
\beq
\label{higgsFCNC3}
{\alpha_3 \over   4 \pi }  {\mu^* \over  m_{\tilde{g}} }\;q_L  Y_d h_u^* d_R   ~.
\eeq
multiplied by a calculable function of ${|m_{\tilde{g}}|\over m_0}$.  Note that $\mu$ and $m_{\tilde{g}}$ ``carry" opposite $R$-charge.  The coupling of the up-type Higgs to the down-type quarks, (\ref{higgsFCNC3}), is of the form expected in a general two-Higgs doublet model.  This leads to large Higgs-mediated FCNCs at large $\tan\beta$, despite the loop suppression factor.

In the MSSM, there are two other classes of diagrams, shown in Fig.~\ref{fig:FCNCtanbeta}, that contribute to couplings like (\ref{higgsFCNC3}). Both diagrams involve a heavy Higgsino in the loop. The first class, due to a Higgsino-squark loop, leads  to an $h_u$ coupling to down quarks with a coefficient  proportional to ${\mu^* A^* \over |m_0|^2} \; Y_u Y_u^\dagger Y_d$ (assuming proportional $A$-terms),   instead of the ${\alpha_3 \mu^* \over m_{\tilde{g}}}$ factor in (\ref{higgsFCNC3}). The second, involving a Higgsino/Wino/Bino-squark loop, is proportional to ${\mu^* m_{\tilde{B},\tilde{W}}^* \over  |m_0|^2} \; Y_d$.  

In the MRSSM, the $PQ$ symmetry acts not only to rotate ($D_R$,$E_R$) but also $R_d$ with a phase opposite that of $H_d$, as required by invariance of the $R$-symmetric $\mu_d$ term (\ref{Wmu}).  Moreover, the $PQ$-symmetry is explicitly broken {\it only} by a dimension-two operator, the $B_\mu$ term (sufficient to avoid an unwanted massless Goldstone boson).  This implies the $PQ$- and $R$- violating couplings (\ref{higgsFCNC1}), the dimension-five $R$-violating gaugino contribution (\ref{higgsFCNC2}), and thus the dangerous couplings (\ref{higgsFCNC3}) are all absent in the $R$-symmetric model.  Moreover, the diagrams involving a Higgsino/Wino/Bino-squark loop also vanish since they involve either $A$-terms, the $R$-violating $\mu$-term, or Majorana masses.  Consequently, these otherwise dangerous contributions to FCNCs at large $\tan\beta$ are absent in the MRSSM.

\section{CP Violation beyond the Flavor Sector}
\label{sec:CPbeyondflavor}

We can count the complex phases of the MRSSM analogously to the counting in the MSSM.  Given completely arbitrary couplings in the superpotential and K\"ahler potential, one performs global phase rotations on the superfields to remove unphysical phases \cite{Dimopoulos:1995ju,Haber:1997if}.

In the flavor-neutral sector of the MRSSM there are a number of complex parameters:  two Higgsino mass terms $\mu_u$ and $\mu_d$; three Dirac gaugino masses $m_i$; three holomorphic scalar masses of the adjoints $M_i^2$; the $B_\mu$ term; and  four Yukawa couplings $\lambda_{u, d}^{\tilde{B}}$, $\lambda_{u, d}^{\tilde{W}}$, totaling 13 complex parameters. There are seven superfields $H_{u, d}, R_{u, d}, \Phi_{\tilde{B},\tilde{W},\tilde{g}}$, whose phases can be used to remove six of the phases from the complex parameters (one irremovable phase corresponds to the unbroken $R$-symmetry).  Note that we chose a basis where the gaugino coupling is real, i.e. we do not allow a rephasing of the gaugino fields.  This implies that the squark and quark fields are rephased as a superfield.  Given this basis, it is easy to see that there are seven complex parameters invariant under rephasings of these seven superfields: $m_i M_i^*$, $i = \tilde{B},\tilde{W},\tilde{g}$, and $\mu_u m_j (\lambda_u^j)^*$, $\mu_d m_j (\lambda_d^j)^*$, $j = \tilde{B},\tilde{W}$.  A priori there is one more phase in the flavor-conserving sector compared to the MSSM \footnote{Without assuming gaugino mass unification in the MSSM or the MRSSM.}.  Now if the Yukawa couplings Eq.~(\ref{Wphi}) were absent (some form of sequestering, for example), there would be only three additional complex parameters $m_i M_i^*$.  This would be reduced to just one complex parameter if gaugino-adjoint mass unification occurred.

\subsection{Constraints from EDMs}

The usual one-loop contributions to EDMs in the MSSM from left-right insertions are completely absent since there is no Majorana gaugino mass nor any left-right squark mass mixing (no $A$-terms or $\mu$-term). The one loop contributions to EDMs with Dirac gauginos in models without an extended $R$-symmetry were  considered in \cite{Hisano:2006mv} and it is easily seen that they all vanish in the $R$-symmetric limit, since they require  either $\mu$, $A$ term or Majorana mass insertions. Two-loop contributions to EDMs from pure gaugino/Higgsino loops are also absent for the same reason. Although we shall see that the electron and neutron EDMs in the present scenario are very small, it is noteworthy that with moderate $R$-symmetry violation (as considered in  \cite{Hisano:2006mv}) contributions arise which are possibly accessible to the next generation of experiments. 

\begin{figure}
\begin{center}
\includegraphics[width=2.2in]{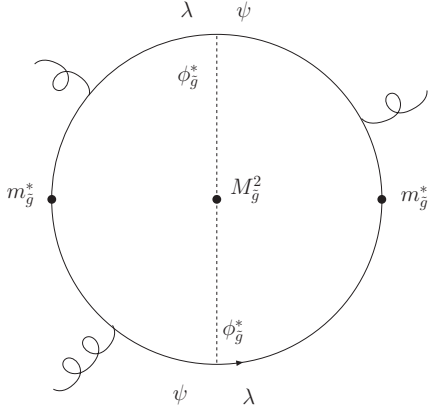}
\end{center}
\caption{The two-loop diagram contributing to the Weinberg operator in the $R$-symmetric model.}
\label{fig:weinberg}
\end{figure}

The leading EDM contribution surviving in the $R$-symmetric model (assuming some mechanism to cancel $\bar{\theta}$) is the one due to the phases in $m_i M_i^*$ and contributes  to the coefficient of  the Weinberg operator, $w G G \tilde{G}$. The contribution to $w$ of the two-loop graph on Fig.~\ref{fig:weinberg}  can be estimated  as $w \sim \frac{\alpha_s^2}{16 \pi^2 |m_{\tilde{g}}|^2} {\rm Arg} (m_{\tilde{g}} M_{\tilde{g}}^*) $, yielding the following 
contribution  to the neutron EDM, see also  \cite{Hisano:2006mv}: 
\beq
\label{neutronEDM}
|d_n| \simeq 4 \times 10^{-26} {\rm  e \; cm} \; \left( {\rm Im} (m_{\tilde{g}} M_{\tilde{g}}^*) \over |m_{\tilde{g}}|^2 \right)  \, \left({1 \; \tev \over |m_{\tilde{g}}|}\right)^2 \, .
\eeq
Thus for TeV-scale masses,  comparison of (\ref{neutronEDM}) to the current upper \cite{Pospelov:2005pr} bound $|d_n| < 6 \times 10^{-26}$ e cm   yields no significant constraint on the phases. 

\begin{figure}
\begin{center}
\includegraphics[width=2.2in]{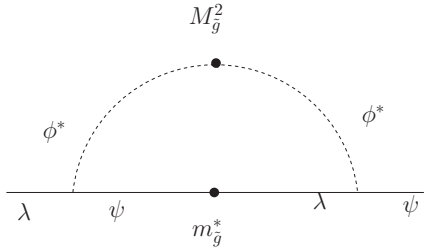}
\end{center}
\caption{The one-loop diagram leading to renormalization of $\bar{\theta}$ in the $R$-symmetric model.}
\label{fig:thetabar}
\end{figure}

\subsection{Constraints from Strong CP}

In the MRSSM the leading order contribution to $\bar{\theta}$ arises from renormalization of the Dirac gaugino
mass at one loop, due to a gaugino/adjoint fermion---scalar adjoint loop, Fig.~\ref{fig:thetabar}, yielding:
\beq
\label{oneloopMD}
\delta m_{\tilde{g}} \sim {\alpha_s \over 4 \pi} {m_{\tilde{g}}^* M_{\tilde{g}}^2 \over |m_{\tilde{g}}|^2}~,
\eeq
and the corresponding contribution to $\bar{\theta}$ is 
\beq
\label{oneloopbartheta}
\delta_{\bar{\theta}} \sim {\alpha_s \over 4 \pi} \; {{\rm Im} (m_{\tilde{g}}^* M_{\tilde{g}})^2 \over |m_{\tilde{g}}|^4}~.
\eeq
Thus, for order one phases,   requiring that $\delta_{\bar{\theta}} < 10^{-9}$, one obtains rather strong constraints on the phases in the gaugino-adjoint sector:  ${\rm Arg}(m_{\tilde{g}}^* M_{\tilde{g}}) \ll 10^{-7}$. 

The constraints from $\bar{\theta}$ on the phases in the squark mass matrices are weaker than in the MSSM. This is because the one-loop squark-gaugino graphs which renormalize the quark masses require Majorana mass and/or A-term insertions, see e.g.~\cite{Hiller:2002um}, and are therefore absent in the $R$-symmetric limit.

\section{Flavor without a continuous $R$-symmetry}
\label{sec:approx}

Until now, we have considered the $R$-symmetry to be an exact continuous symmetry.
Virtually all of the benefits to low energy supersymmetry that we have described 
are maintained if only a (large enough) discrete subgroup of the $R$-symmetry is preserved.  
For instance, if a $Z_6$ subgroup of the continuous $R$-symmetry is preserved, then all 
that has been described here is still applicable.  That is, all the operators in question ($A$-terms, dimension five proton decay operators, Majorana gaugino masses 
and the $\mu$ term) remain forbidden.  Even if the subgroup is just $Z_4$, all of the above
benefits apply, with the exception that  dimension-5 proton decay operators are now allowed.

\subsection{Modifications from a weak breaking to $R$-parity}
Even if there is weak breaking of the $R$-symmetry to a $Z_2$ (i.e., $R$-parity), the larger $R$-symmetric contributions can still serve to reduce the extent of 
the supersymmetric flavor problem. One possible weak breaking of the $R$-symmetry can arise from the conformal anomaly.  This causes a Majorana mass for the gauginos, generally expected to be of order $\delta M \sim \frac{m_{3/2}}{16 \pi^2}$.  However, since we still have sizeable Dirac gauginos masses, it is interesting to consider the effects of suppressed Majorana masses on top of this.  In addition it is possible that a $\mu$-term or $A$-terms could also be generated from $R$-symmetry breaking, leading to left-right mixing.  

\subsubsection{Corrections to $\Delta F=2$}
In the presence of a small Majorana mass $\delta M$ to the gluino, there are additional contributions to $K$-$\bar{K}$ mixing:
\be
\Delta M_{\rm box} = 2 (\delta C_1 M_1 +\delta C_2 M_2 + \delta C_3 M_3 + \delta C_4 M_4 +\delta  C_5 M_5),
\ee
where $M_{1,4,5}$ are as before, see Eq.~(\ref{m145}), and:
\bea
M_2 &=&-\frac{5}{24} \left( \frac{m_K}{m_s+m_d} \right)^2 m_K f_K^2 B_2, \nonumber \\
M_3 &=&\frac{1}{24} \left( \frac{m_K}{m_s+m_d} \right)^2 m_K f_K^2 B_3.
\eea
The corrections to the coefficients are:
\bea
\delta C_1 &=& \frac{\alpha_s^2}{216 m_{\tilde q}^2} \frac{ 24 x \delta M^2}{m^2_{\tilde g}} f_6(x)( \delta_{LL}^2 + \delta_{RR}^2)~, \nonumber \\
\delta C_2 &=& \frac{\alpha_s^2}{216 m_{\tilde q}^2}  \frac{204 x \delta M^2}{m^2_{\tilde g}} f_6(x)( \delta_{LR}^2 + \delta_{RL}^2)~,\nonumber \\
\delta C_3 &=& \frac{\alpha_s^2}{216 m_{\tilde q}^2}  \frac{-36 x\delta M^2}{m^2_{\tilde g}} f_6(x)( \delta_{LR}^2 + \delta_{RL}^2)~, \\
\delta C_4 &=&  \frac{\alpha_s^2}{216 m_{\tilde q}^2}(132 \delta_{LR} \delta_{RL} \tilde f_6(x) \nonumber \\ &&+ \frac{504 x\delta M^2}{m^2_{\tilde g}} f_6(x) \delta_{LL} \delta_{RR})~, \nonumber\\
\delta C_5 &=&  \frac{\alpha_s^2}{216 m_{\tilde q}^2}(-180 \delta_{LR} \delta_{RL}+\frac{24 x \delta M^2} {m^2_{\tilde g}}f_6(x) \delta_{LL} \delta_{RR} )~.\nonumber
\eea
Here again $x = m_{\tilde g}^2/m_{\tilde q}^2$, $B_{2,3} = 0.66,1.05$ are bag factors for the additional operators \cite{Ciuchini:1998ix}.  By explicit calculation, one can determine that for $\delta M^2/m^2_{\tilde g}\lsim 10^{-2}$, the Majorana contributions are subdominant. Since, for $m_{3/2} \sim m_{\tilde g}$ we expect this ratio to be $O(10^{-4})$, and these contributions should not, in general, be important.

Neglecting the contributions from Majorana insertions, but retaining the left-right insertions, we obtain constraints on the size of the left-right mixing, shown in Fig.~\ref{fig:lr}(a).  A similar calculation can be done for $B$ meson mixing, and the constraints we obtain are shown in Fig.~\ref{fig:lr}(b).  Satisfying the bounds on the CP-violating $\epsilon_K$, however, is more difficult without more squark degeneracy or smaller CP-violating phases.  Analogously, it is also considerably more difficult to satisfy the constraint from $\epsilon'/\epsilon$ when left-right mixing is present.

\subsubsection{Corrections to $\Delta F=1$}
\begin{figure*}
a)\includegraphics[width=2.2in]{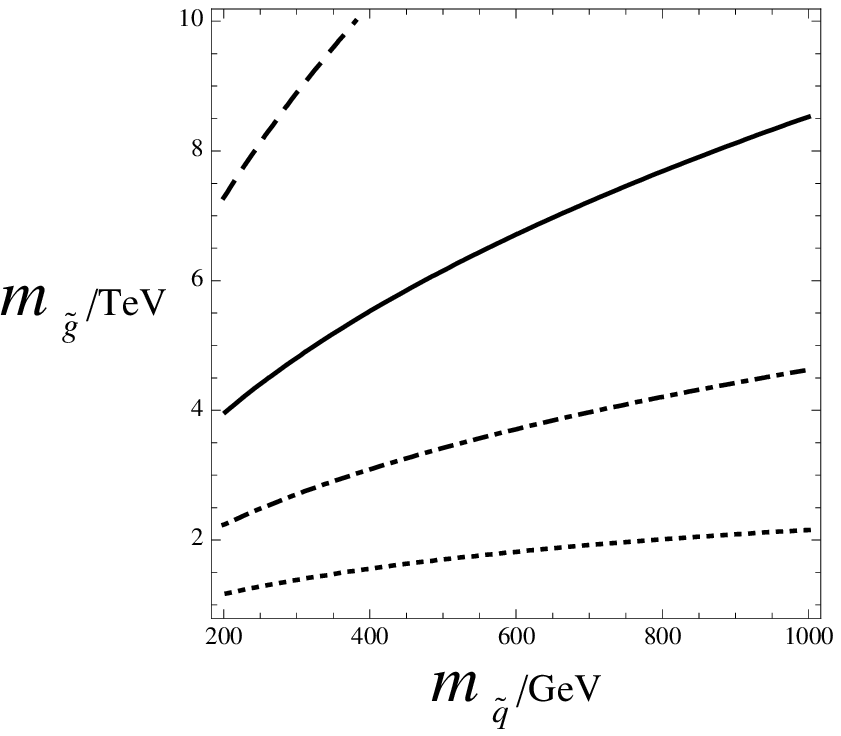}
b)\includegraphics[width=2.2in]{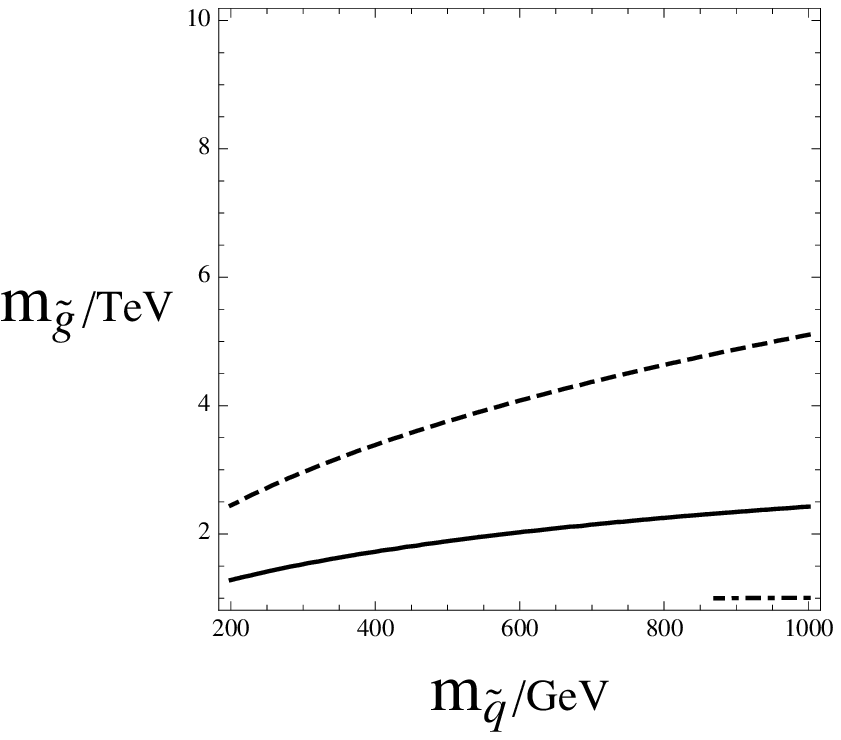}
c)\includegraphics[width=2.2in]{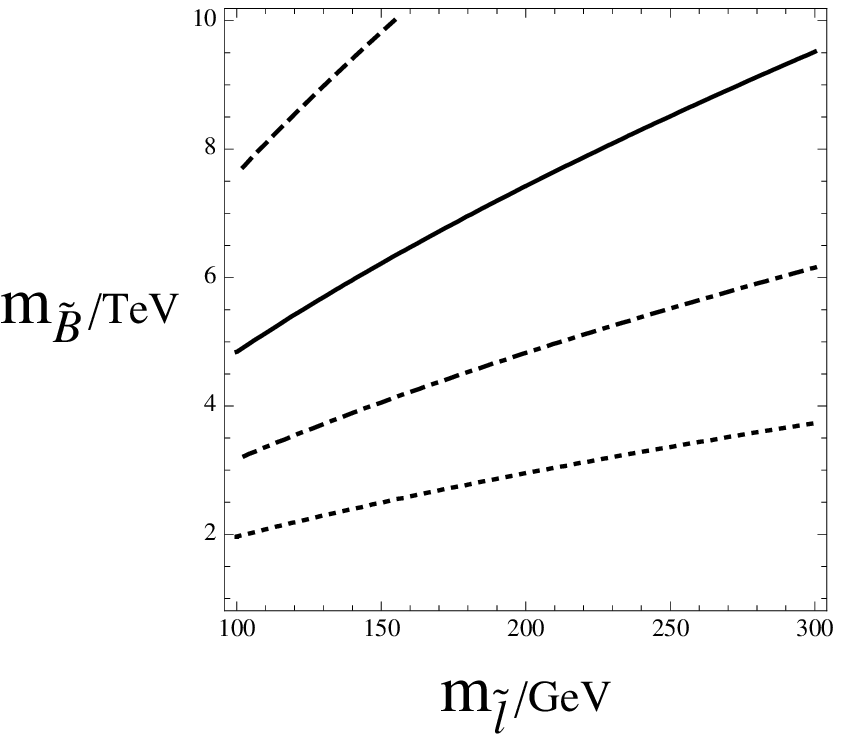}
\caption{Contours of the maximum flavor-violating left-right insertion
where 
(a) $\Delta M_{\rm box} = \Delta M_k$ for $K$-$\bar{K}$ mixing
(as in Fig.~\ref{fig:kkbarlimits});
(b) $\Delta M_{\rm box} = \Delta M_{B_d}$ for $B$ meson mixing
(as in Fig.~\ref{fig:bmix});
(c) $BR_{\mu \rightarrow e \gamma} =1.2 \times 10^{-11}$ 
for $\mu \rightarrow e\gamma$
(as in Fig.~\ref{fig:muegamma}), with $\delta M/m_{\tilde{B}} = 10^{-2}$.
In each case, we took 
$\delta_{LR} = \delta_{RL}$ and $\delta_{LL}=\delta_{RR}=0$
for the flavor-mixing entries in the relevant squark or
slepton mass matrix.
Contours are $\delta =0.03,0.1,0.3,1$ for dotted, dot-dashed, 
solid, dashed respectively.}
\label{fig:lr}
\end{figure*}

The strongest constraint from lepton flavor violation  is from $\mu \rightarrow e\gamma$. 
It is interesting to note that even in the presence of left-right mixing, there is suppression of flavor violation given 
mostly Dirac gauginos. The contribution to $\mu \rightarrow e \gamma$ from left-right insertions is \cite{Hagelin:1992tc}:
\bea
\label{muegammaLR}
 BR_{\mu \rightarrow e \gamma} = \frac{48 \alpha \pi^3}{G_F^2} \; |A_{LR}|^2~,
  \eea
with: 
\bea
A_{LR} = {\alpha_Y \over 4 \pi} {\delta_{LR} \tilde{m}_R^2 \over \tilde{m}_L^2 - \tilde{m}_R^2} {\delta M \over m_\mu} \;
\left({f_{3n}(x_R)\over \tilde{m}_R^2} - {f_{3n}(x_L) \over \tilde{m}_L^2} \right),
\eea
where $x_{L(R)} = m_{\tilde{B}}^2/\tilde{m}_{L(R)}^2$ and $$f_{3n}(x) = \frac{1 + 2 x \log x - x^2}{2 (1-x)^3}.$$ In Fig.~\ref{fig:lr}(c) we show the bounds on the Bino and slepton ($m_{\tilde{l}}= \tilde{m}_L = \tilde{m}_R$) masses, for various values of $\delta_{LR}$, assuming that the contribution of (\ref{muegammaLR}) alone is less than the observed value and that the ratio of Majorana to Dirac mass is $\delta M/m_{\tilde{B}} = 10^{-2}$.

The contribution to the  $b \rightarrow s \gamma$ branching ratio 
from left-right insertions is \cite{Gabbiani:1988rb,Gabbiani:1996hi}:
\bea
&&BR_{b\rightarrow s \gamma}=\nonumber \\
&& \frac{\alpha_s^2 \alpha m_b^3  \tau_B}{81 \pi^2 m_{\tilde q}^4} \; 
  \left|  m_b \delta M {G_{n2}(x)\over 2} \delta_{LR} \right|^2 + L\leftrightarrow R ~,
\eea
where $G_{n2}$ is defined in (\ref{gfunctions}).
Note that this contribution can enter only with both a left-right insertion
as well as a Majorana gaugino mass.  Nevertheless, this is still well below the observed value. 
As before, taking $m_{\tilde g}=1.5$~TeV, $m_{\tilde q}=300$~GeV,  we find:
\bea
&& \delta BR_{b\rightarrow s \gamma} = \nonumber \\
&&\left(1.75 \times 10^{-7} \delta_{LR} \delta_{LL} \frac{\delta M}{10 \; \gev} \right. \nonumber \\
 &&+ \left. 5.2 \times 10^{-7} \delta_{LR}^2 \left(\frac{\delta M}{10 \; \gev} \right)^2 \right) + (L \leftrightarrow R).
 \eea
which is well below the bound (\ref{eq:bsgbound}).

\section{UV Realization and Unification}
\label{sec:models}
 
\subsection{Unification}

The presence of the additional adjoint states and/or Higgs states in the theory raises the question of perturbative unification. Two groups can contain these adjoints and still plausibly unify perturbatively, namely ``trinification'' ($SU(3)^3$, see e.g.~\cite{Fox:2002bu}) or $SU(5)$. In each of these cases, we can complete the new fields to unified multiplets by adding ``bachelor'' fields.  In the case of $SU(3)^3$ this amounts to the addition of a vectorlike pair of fields $({\bf 1, 2, \pm 1/2})$,  two pairs of $({\bf 1, 1, \pm 1})$, as well as   four singlets. In $SU(5)$, we must add $({\bf 3,2,-5/6})$ and $({\bf \bar 3,2,5/6})$ \cite{Fox:2002bu}. The GUT-completed adjoint amounts to three and five additional flavors in $SU(3)^3$ and $SU(5)$, respectively. 

In the case of $SU(3)^3$ we can identify the new $({\bf 1, 2, \pm 1/2})$ fields as the two Higgs doublets, with zero $R$-charges, consistent with those of the $SU(3)^3$ adjoint fields. The fields $R_{u,d}$ we include as a split multiplet. The additional pair of $({\bf 1, 1, \pm 1})$ fields will not acquire mass unless  $SU(3)^3$ is broken, so a combination of the $R$-symmetry generator and the GUT symmetry generator is preserved, such that their  
$R$-charges are $2, 0,0,-2$. This then allows for supersymmetric mass terms of the charged ``bachelors" which preserve a $Z_4$ subgroup of the $R$-symmetry.

 In the case of $SU(5)$, it is quite difficult to arrange the GUT and $R$-symmetry breaking such that the bachelor fields have $R$-charges allowing $Z_4 \subset U(1)_R$-symmetric mass terms. Thus, we must invoke a small $R$-symmetry breaking to give these fields a mass.  Another alternative would be the one employed by \cite{Chacko:2004mi}, in which the adjoint fields themselves are composite, and thus did not contribute to the running of the gauge couplings above a TeV.

Ultimately, our main focus here is on the flavor properties of this theory. GUT model building is a subtle and worthwhile question which we defer to future work.

\subsection{UV Completion}
\label{sec:UVcompletion}

There are several issues which arise when embedding the low energy effective model
into a UV completion.  One issue is the possible linear potential term for the singlet (the $U(1)_Y$  ``adjoint"), which is known to lead to
a destabilizing divergence \cite{Bagger:1995ay}.  Another issue is suppressing the kinetic mixing 
between $U(1)'$ and hypercharge.  

One resolution of these problems can be accomplished by having a naturally low cutoff
scale for all of the higher dimensional operators.  For example, consider a two-brane 
RS1-like setup \cite{Randall:1999ee}. In the bulk we add the   vector and adjoint  
superfields (as $N=2$ partners of the gauge fields) and Higgs fields, while on the IR brane we put the 
matter fields and supersymmetry breaking, which we implement using a single superfield 
$X$ of $R$-charge 2, with a linear superpotential:
\be
W \supset \mu^2 X , 
\ee
and additional terms:
\be
\int d^4 \theta \; \frac{(X^\dagger X)^2}{\Lambda^2}
\ee
that stabilize the scalar component of $X$ at the origin.
Thus $X$ acquires an $F$-component expectation $\vev{X} = \theta^2 \mu^2$ and supersymmetry is broken while $R$-symmetry is preserved. Because the strong coupling scale and the SUSY breaking scale are assumed to be comparable, higher dimensional operators will be very important. 
In particular,  the field combination:
\be
\bar{D}^2 D_\alpha \frac{X^\dagger X}{\Lambda^2},
\ee
where $\Lambda \sim \mu$ is the IR strong-coupling scale, has the same structure and $R$-charge as a field strength of a $U(1)'$ acquiring a $D$-term, but does not actually correspond to a genuine $U(1)'$. Consequently, the issue of $U(1)'$-hypercharge mixing is moot.  Similarly, there is no concern of generating large $D$-terms, because they are just a recasting of the $F$-terms.
Moreover, the radiative corrections to the linear potential generated by supergravity for the 
singlet are cut off at the scale $\Lambda$ and are thus safe.

In addition, as in other models with similar structure, IR contributions to the Higgs potential 
give a possible large contribution to the quartic:
\be
\int d^4 \theta \frac{X^\dagger X}{\Lambda^4}  ((H_u^\dagger H_u)^2 + H_u^\dagger H_u H_d^\dagger H_d+(H_d^\dagger H_d)^2),
\ee
thus the theory can reasonably exist at large or small $\tan \beta$.
In this scenario the SUSY breaking scale is small, with a gravitino mass of order $\tev^2/\mpl$, 
leading to phenomenology similar to gauge mediation.

Finally, another UV issue is understanding the (little) hierarchy between the Dirac gaugino masses and the soft scalar masses.  One interesting possibility is to separate the matter sector from the hidden sector across an extra dimension, with gaugino fields in the bulk. The gauginos are thus able to pick up SUSY breaking directly from the hidden sector while the scalar masses receive only subdominant contributions from bulk-field mediation. Ordinarily, with only gauge fields in the bulk, this leads to gaugino mediation \cite{Kaplan:1999ac,Chacko:1999mi}. With additional light bulk states, however, the usual flavor-blind sequestering may not be effective \cite{Anisimov:2001zz}, leading to effective operators communicating supersymmetry breaking to the matter sector that violate flavor but may nevertheless be volume suppressed compared with the Dirac gaugino mass operators.  Pursuing a more detailed model would be very interesting, but we leave it for future work.

\section{Phenomenology}
\label{sec:pheno}

There are several novel phenomenological features of our $R$-symmetric 
model.  The most unusual characteristic is that 
large flavor violation is allowed in the squark and slepton mass matrices.  
The presence of large flavor violation in this theory means that it is no longer appropriate to discuss ``stops'' or ``selectrons'' necessarily, as we do not expect a strong alignment between the superpartner mass basis and the Yukawa basis. This large flavor violation can lead to interesting consequences, such as bizarre cascades where squarks decay into other squarks, single production of squarks which decay into $b$-jets and missing energy (we do not say ``bottom squarks'' for the aforementioned reason). 
For instance, at the LHC a single top final state can arise from ordinary di-squark production, 
$pp \rightarrow \tilde{q}\tilde{q}^*$, after one squark decays
into a top while the other into a light quark flavor. We also note that other supsersymmetric scenarios with sizeable flavor violation have been recently considered \cite{Feng:2007ke,Nomura:2007ap}.

Slepton flavor violation also provides a powerful method
to probe the flavor structure of the slepton mass matrices.
For instance, slepton production and decay into $\ell_i\ell_j$ final states
($i \not= j$) provides a window to study the mass matrix 
structure at the LHC \cite{Bityukov:1997ck,Agashe:1999bm,Bartl:2005yy}.
This would be a bonanza for a future linear collider
(for example \cite{ArkaniHamed:1996au}).
In addition, the large CP phases in the slepton mass matrices
can also be probed at the LHC through slepton CP asymmetries.
As emphasized in \cite{ArkaniHamed:1997km}, colliders are sensitive 
to the rephase invariant 
$\tilde{J} \propto {\rm Im}(m_{12}^2m_{23}^2m_{31}^2)$, 
which is essentially unconstrained by charged lepton flavor violation and EDMs.
Additionally one might consider looking for endpoints in $\mu$-$e$ invariant masses. The large flavor violation that is expected here opens up the possibility for a wide variety of new signals at the LHC and is worthy of significant study.

Dirac gauginos also provide a rich phenomenology \cite{Fox:2002bu}. However, as these gauginos are likely quite heavy, single production may be the only way they will be seen on shell. If the $R$-symmetry is very good (and so Majorana masses are small or absent), then we expect no like-sign dilepton signatures at the LHC.

There are various other issues that we should also mention.
One obvious concern is that the $SU(2)$ triplet acquires a vev, which can yield a dangerous correction to the $\rho$ parameter. The vev of the real part of the triplet is found to be:
\be
\vev{\phi_{\tilde W}} = \frac{\sqrt{2} g_2 v^2 m_{\tilde{W}} \cos 2 \beta }{8 m_{\tilde{W}}^2 + \delta^2}
\ee
where $\delta^2$ is the sum of all of corrections to the electroweak triplet scalar mass squared beyond those in (\ref{diracmassoperator}), for example, from the soft terms (\ref{adjsoft}). Taking the case of $\delta^2 =0$, one finds 
$|\Delta \rho| \simeq 2 \vev{\phi_{\tilde W}} ^2/v^2 = 
g_2^2 v^2 \cos^2 2 \beta/ 16 m_{\tilde{W}}^2 \approx 8 \times 10^{-4} \cos^2 2 \beta \,\, (1 \,\, \tev/m_{\tilde{W}})^2$. 
Thus, for Winos $m_{\tilde{W}} \gsim 1$~TeV we are consistent with precision electroweak limits.

As is typical for Dirac gauginos, the presence of the operator in Eq.~(\ref{diracmassoperator}) cancels off the tree-level Higgs quartic from the SUSY $D$-terms. Since this quartic is important for generating the mass of the Higgs, we remind the reader of the possible solutions, as described in \cite{Fox:2002bu}. The simplest possibility is the inclusion of a term:
\be
W \supset S H_u H_d
\ee
in the theory, such as in the NMSSM.  This term generates a potentially large quartic for the Higgs at small $\tan \beta$. Because of the additional matter, RG running can make this larger than in the NMSSM \cite{Masip:1998jc,Espinosa:1998re}. 

Alternatively, we can include additional scalar masses for the $SU(2)$- and $U(1)_Y-$ adjoints, such that integrating them out does not kill off the quartic. Unlike in \cite{Fox:2002bu} such terms would be natural here, as we are including $F$-term $R$-symmetric SUSY breaking.

Finally, one can simply allow the quartic to be strongly suppressed, and have the dominant contribution to the quartic generated from radiative corrections from the scalar tops. However, this will require heavy ($\sim\tev$) stops, which will make the theory more tuned.

Because the theory has heavy gauginos, the Bino is no longer a dark matter candidate. If the SUSY breaking is small, and the gravitino is light, one must appeal to a new symmetry and fields (like messenger parity \cite{Dimopoulos:1996gy}) or an axion. However, it is also interesting to understand what  the dark matter candidates are if the SUSY breaking scale is high and the mediator is gravity. In this case, the Higgsino can still be the LSP, or, more simply, in the case that the NMSSM-like mechanism is employed to generate a quartic at small $\tan \beta$, Higgsino-singlino mixing is expected after electroweak symmetry breaking, and thus a mixed singlino-Higgsino is a viable and natural dark matter candidate \cite{Hsieh:2007wq}.

\newpage
\section{Conclusions}
\label{sec:conc}

We have presented a new and radically different approach to the supersymmetric flavor problem, based upon the presence of a continuous or extended discrete $R$-symmetry in the low energy theory. This approach allows large flavor violating masses, even with light sfermions, and is consistent with present precision measurements.

We have constructed the minimal supersymmetric standard model with such an $R$-symmetry.  The MRSSM has Dirac gaugino masses, an extended Higgs sector, no $A$-terms, and no left-right squark or slepton mass mixing.  We have calculated the consequences of these modifications for flavor violating observables and find that in natural regions of parameter space, where gauginos have masses $O({\rm TeV})$ and sfermions are in the $200$-$500$~GeV range, $O(1)$ flavor violation is consistent with present observations. We thus argue that $R$-symmetric supersymmetry is a natural solution to the supersymmetric flavor problem, and can be naturally embedded within a gravity-mediated framework.

The MRSSM has dramatically different phenomenology from the MSSM. As is typical for Dirac gauginos, there is a moderate hierarchy between lighter scalar and heavier gaugino masses. Additionally, these theories have copious flavor violating signals, which are typically taken for granted to be small in other extensions of the standard model. Precision studies of B physics or improvements in tests of $\mu \rightarrow e \gamma$ may probe this in the near future. It is possible that such flavor violation may be visible at the LHC.

A tremendous amount of work remains to be done. QCD corrections to the meson mixing operators may be significant, as in the MSSM. Global fits to the precision flavor observables should place more stringent constraints on flavor violation in the squark sector. The phenomenology of dark matter must be explored. However, it is remarkable that such a dramatically different solution to the flavor problem exists, and lends credence to the idea that a wide variety of unexpected signals may await us at the LHC.

\vskip 0.15in
\onecolumngrid
\section*{Acknowledgments} 

The authors thank CERN and the organizers of the CERN-TH Beyond the Standard Model Institute, 
and the French Alps, where this work was initiated. The authors thank 
A.~Arvanitaki, 
A.~Blechman, 
P.J.~Fox,
R.~Harnik,
D.E.~Kaplan, 
Z.~Ligeti,
T.~Roy,
N.~Toro,
and J.~Wacker, 
for stimulating and useful discussions, as well as A.E.~Nelson for reading a draft of the paper and providing constructive comments. NW thanks S.~Dimopoulos and the SITP at Stanford as well as M.~Peskin and the SLAC theory group for their hospitality and support while this work was being undertaken, and the IFT-UAM/CIS for hospitality and support under Proyecto de la Comunidad de Madrid HEPHACOS S-0505/ESP-0346 as this work was completed.
GDK was supported by the Department of Energy under grant number DE-FG02-96ER40969.
EP was supported by NSERC.  
NW was supported by NSF CAREER grant PHY-0449818 and DOE grant number 
DE-FG02-06ER41417.

\bibliography{mrssm}

\bibliographystyle{apsrev}

\end{document}